\documentclass{aa}
\usepackage{graphics,epsfig,amsmath,amssymb,amstext}

\begin{document}

\thesaurus{1(02.01.1; 02.14.1; 09.03.2; 10.01.1)}

\title{Bimodal production of Be and B in the early Galaxy}

\author{Etienne Parizot \and Luke Drury}


\institute{Dublin Institute for Advanced Studies, 5 Merrion Square,
Dublin 2, Ireland\\ e-mail: parizot@cp.dias.ie; ld@cp.dias.ie}

\date{Received date; accepted date}

\maketitle

\begin{abstract}

Recently models based on the acceleration of metal-rich material
inside superbubbles have been proposed to account for the observed
abundances of Be and B in metal-poor halo stars.  We analyse some of
the implications of these models for the distribution of the Be/O and
B/O abundance ratios.  In particular, we discuss the possible scatter
in the data and argue that LiBeB production in the very early Galaxy
was probably bimodal, with isolated supernovae giving rise to a
low-efficiency mechanism, and collective supernovae exploding in an OB
association inducing a high-efficiency mechanism.  This should produce
two populations of halo stars, one with high L/M ratios (light
elements/metals), and the other with L/M ratios about ten times
lower. The relative weight of these two populations depends on the
fraction of supernov\ae~exploding inside superbubbles.  In this
context, we discuss the recent observation of the B-depleted,
Li-normal star HD 160617 (Primas, et al., 1998), as well as the
reported spread in the Be data at [Fe/H] $\sim -2.2$ (Boesgaard, et
al., 1999b).  Finally, we predict that Be will be found to be even
more deficient than B in HD 160617.


\end{abstract}

\section{Introduction}

In the last decade, the high sensitivity of the KECK telescope and the
HST has allowed numerous observations of the Be and B abundances in
halo stars having very low metallicities, down to $[\mathrm{Fe/H}] =
-3$, i.e. $\mathrm{Fe/H}\sim 10^{-3} (\mathrm{Fe/H})_{\odot}$ (e.g.
Molaro, et al., 1997; Duncan, et al., 1997; Garcia-Lopez, et al.,
1998).  These observations show a clear proportionality between the
Be, B and Fe abundances.  Considering that the light elements are
secondary nuclei synthesized by spallation from C and O nuclei
(Reeves, et al., 1970), it had been expected instead that their
abundance would increase as the square of the metallicity
(Vangioni-Flam, et al., 1990).  This \emph{secondary behavior} follows
directly from the standard Galactic Cosmic Ray Nucleosynthesis (GCRN)
scenario (Meneguzzi, et al., 1971), in which most of the Be and B are
produced by energetic protons and $\alpha$ particles interacting with
C and O nuclei accumulated in the ISM (direct spallation).

The most natural way to account for the unexpected constancy of the
Be/Fe and B/Fe ratios is to assume that Be and B nuclei are mainly
produced by \emph{reverse spallation}, i.e. by energetic C and O
nuclei accelerated shortly after their release into the interstellar
medium (ISM), and interacting with ambient H and He nuclei (Duncan, et
al., 1992; Cass\'e, et al., 1995).  This makes the production rates
independent of the ambient metallicity, and the amount of light
elements (L elements) in the Galaxy therefore increases jointly with
the most abundant metals (M elements), namely C, O and Fe.  This
behavior is refered to as \emph{primary}, and has been shown to follow
naturally from the assumption that most of the supernova (SN)
explosions occur in OB associations.  This is the heart of the
so-called superbubble models, whose main lines we recall in
Sect.~\ref{SBmodels}.

Some recent observations have suggested that [O/Fe] continues to
decrease at the lowest metallicities, rather than reaching a "plateau"
value which is the same for all metal-poor stars.  These observations
are still controversial (Fulbright \& Kraft, 1999), although they
could be accounted for by allowing for different `mixing times' for
the freshly ejected Fe and O nuclei in the ISM (Ramaty, et al., 1999). 
They have raised the question whether a primary process for Be and B
production in the early Galaxy is still needed (Fields \& Olive,
1999).  According to both energetics considerations (Ramaty, et al.,
1999; Parizot \& Drury, 2000) and a detailed analysis of the available
data (Fields, et al., 2000), it is now widely agreed that the answer
is yes, at least at a metallicity lower than a so-called transition
metallicity, $Z_{\mathrm{t}}$, say for $\log(Z_{\mathrm{t}}/Z_{\odot})
\equiv [\mathrm{O}/\mathrm{H}]_{\mathrm{t}} \la -1.5$.  In this paper,
we concentrate on the very early Galaxy, when the primary behavior
dominates, and discuss the implications of the superbubble model for
the distribution of the (light elements)/(metals) ratios (namely Be/O
and B/O, or L/M for short) in very metal-poor stars.

\begin{figure}
\centerline{\psfig{file=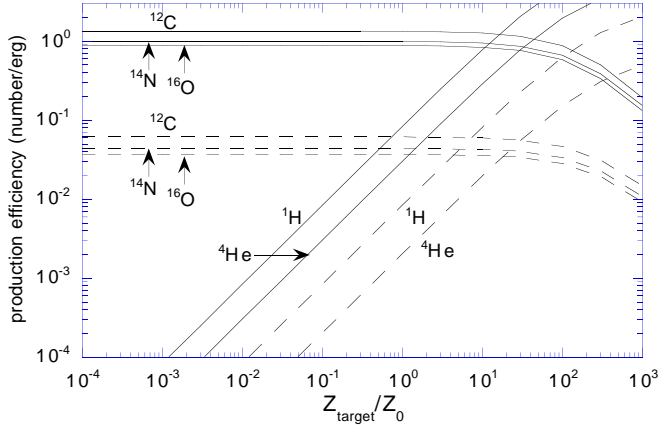, width=\columnwidth}}
\caption{Be production efficiency by spallation for different
projectiles, in numbers of Be nuclei produced per erg of EPs injected,
as a function of the ISM (target) metallicity.  Plain: SB spectrum.
Dashed: CRS spectrum.}
\label{BeProdEff}
\end{figure}

\section{The superbubble models}
\label{SBmodels}

In the very early Galaxy, the interaction of energetic particles (EPs)
having the ISM composition would produce very little Be and B, because
the C and O nuclei are so rare.  Simple energetics considerations thus
indicate that Be and B can be significantly produced only if the EPs
have a composition richer in C and O than the global ISM. This is the
reason why the only viable models proposed so far involve the
acceleration of C- and O-rich material inside a superbubble (Higdon,
et al., 1998; Parizot \& Drury, 1999c,2000, Bykov, 1999; Ramaty et
al., 1999; Bykov., et al., 2000).  Indeed, repeated SNe occurring in
an OB association are known to generate a superbubble (SB) with a
typical radius of order 300~pc, filled with hot, tenuous gas, and
composed of the metal-rich ejecta of the SNe having already exploded,
possibly diluted by the swept-up ambient material (of essentially zero
metallicity in the early Galaxy).

As discussed in detail in Parizot \& Drury (1999c), the SB models for
Be and B Galactic evolution are based on the following sequence of
events: 1) CNO nuclei are ejected by SNe inside the superbubble; 2)
the CNO nuclei are mixed with some ambient, metal-poor material; 3)
the resulting material (including CNO) is accelerated; 4) LiBeB is
produced by spallation through the interaction of these `superbubble
energetic particles' (SBEPs) with the metal-poor material in the
supershell and at the surface of the adjacent molecular cloud; 5) the
LiBeB produced is mixed with the CNO ejected by the SNe, which leads
to a unique value of the L/M ratios throughout the superbubble (and
part of the supershell); 6) new stars form by condensation of this
gas, after possible dilution by ambient, metal-poor gas (from the
supershell or the adjacent molecular cloud.  All these new stars then
have the same L/M ratios, but possibly different overall metallicity.

Apart from this common `astrophysical background', the models proposed
differ in some of their assumptions, notably relating to the
composition and spectrum of the metal-rich EPs.  Ramaty et al.  note
that the current composition and spectrum of the cosmic rays (CRs)
provide a Be production efficiency sufficient to explain the high L/M
ratios observed in halo stars.  This is reminiscent of the original
result of Meneguzzi et al.  (1971), which is the heart of the GCRN
scenario for light element production (Vangioni-Flam, et al., 1990;
Fields \& Olive, 1999): multiplying the light element production rates
from GCRs by the age of the Galaxy, one obtains approximately the
total amount of Be and B present today in the Galaxy.  However, while
the GCRN scenario assumes that the CR composition follows that of the
ISM (i.e. is richer and richer in C and O) and therefore does not
reproduce the primary behavior of Be and B in the early Galaxy, Ramaty
et al.  assume that the CR composition does not change during the
whole Galactic evolution.  This is indeed expected if the CRs are made
of SN ejecta accelerated inside a superbubble, by the shock of
subsequent SNe.  Their composition is then almost independent of the
ISM metallicity, provided that the SN ejecta are not well mixed with
the ambient matter before the acceleration occurs.

In our model (Parizot \& Drury, 1999c), we argue that an acceleration
mechanism different from the diffusive shock acceleration could occur
inside SBs, because of the specific physical conditions prevailing
there (hot, tenuous gas, strong magnetic turbulence, multiple weak
shocks\ldots).  Such a mechanism has been described by Bykov
(1995,1999) and leads to a different energy spectrum, which we refer
to as the `SB spectrum', and which is flatter than the cosmic-ray
source spectrum (CRS) at low energy, say below a few hundreds of MeV/n
(for a discussion, see Parizot \& Drury, 2000).  The actual shape of
the spectrum above this `break' (whether a steep power-law or the
standard CRS shape in $p^{-2}$) is irrelevant here, since it does not
affect the Be and B production efficiency.  We adopt the following
shape for the SB spectrum: $Q(E) \propto E^{-1}$ up to
$E_{\mathrm{break}} = 500$ MeV/n, and $Q(E)\propto E^{-2}$ above.

As can be seen from Fig.~\ref{BeProdEff}, this spectrum makes the
light element production more efficient than the standard CRS
spectrum, so that the same amount of Be and B can be produced by less
metal-rich EPs.  In particular, the observed L/M ratios in halo stars
can still be accounted for if one allows for a perfect mixing of the
SN ejecta with the metal-free material swept-up and evaporated off the
supershell, before the acceleration occurs (Parizot \& Drury,
1999c,2000).  This is suggested by the comparison between the mixing
time inside a superbubble ($\sim 10^{6}$ yr; see Parizot \& Drury
1999c) and the typical age of a superbubble ($\sim 3\,10^{7}$ yr).  In
the following, we analyse the common and distinctive implications of
the above models for the distribution of the L/M ratios in halo stars.

\section{Intrinsic scatter in the L/M ratios}
\label{scatter}

As recalled above, the constancy of the L/M ratios in the framework of
SB models relies on the mixing of the primary SN ejecta with the
secondary light elements produced by the SBEPs, before the formation
of a new generation of stars.  In practice, however, such a mixing
cannot be perfect and the value of the local L/M ratio is expected to
vary from one place to another.  In addition, the formation of new
stars can occur before all the massive stars explode and/or the
induced LiBeB production occurs.  As a result, stars with somewhat
different L/M ratios should form from a given superbubble, and this
should be observed as a scatter in the Be and B data.  This is a
common prediction of any SB model.  However, quantitatively, the
amplitude of the scatter depends on the mixing of the gas inside the
SB, and of the SB gas with the ambient supershell.  Therefore, a model
which assumes that the SN ejecta are well mixed with the ISM
evaporated inside the SB (Parizot \& Drury) predicts a smaller
dispersion than a model in which the SBEPs are almost pure ejecta
(Ramaty et al.), not diluted with ambient gas.  The exact distribution
of the L/M ratios expected in the framework ot these two models is not
calculated in this paper, because it depends on the details of the gas
dynamics inside the SB and the surrounding shell, as well as on the
star formation dynamics.  Instead, we argue that the accumulation of
Be, B and O data could optimistically provide an interesting way to
constrain the models, through the statistical description of the
scatter in the elemental ratios.

In fact, one might already conclude from the very existence of a well
defined correlation between Be and O (or B and O) that the secondary
light elements must be quite well mixed with the primary ejecta (CNO)
before new stars form.  This is an argument in favour of our model,
because a good mixing between the CNO inside the SB and the LiBeB
produced in the SB shell first requires a good mixing of the gas
inside the SB itself.  However, stronger conclusions cannot be drawn
until more data are available and the error bars become small enough
to allow for a direct measure of the scatter in the L/M ratios.  It is
worth noting also that a larger dispersion should be found for Be than
for B, since part of the boron is expected to be produced along with C
and O in the course of SN explosions (by neutrino-spallation; Woosley,
et al., 1990), and thus be `ready-mixed' inside the SBs.

The above source of scatter in the L/M ratios is inherent in the SB
models.  It results from the fact that the light elements are produced
in a different place from CNO, namely in the shell rather than inside
the SB, where the gas in well mixed.  In the following section, we
discuss another source of dispersion in the L/M ratios of
low-metallicity stars, resulting from the fact that SBs do not cover
the whole volume of the Galaxy and therefore an other Be and B
production mechanism dominates in the regions distant from SBs.

\section{Bi-modal LiBeB production}

\begin{figure}
\centerline{\psfig{file=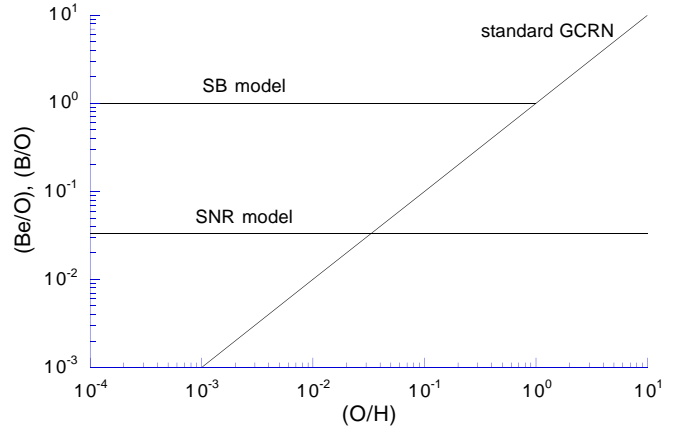, width=\columnwidth}}
\caption{Schematic view of the L/M production ratios as a function of
the ISM metallicity, as expected from the GCRN scenario, the SB model
and the SNR model (Parizot \& Drury, 1999a,b).  The abscissa is
normalized to the metallicity when GCRN dominates the LiBeB
production, i.e. approximately between $[\mathrm{O}/\mathrm{H}] =
-1.5$ and $-2$ (Fields, et al., 2000; Parizot \& Drury, 2000), and the
ordinate to the L/M ratios at this time.  The intrinsic scatter around
each line is not shown.}
\label{L/M}
\end{figure}

The vast majority of spallogenic Li, Be and B nuclei are produced by
particles of relatively low-energy, which are just the most numerous.
Now since only the SBEPs of highest energy can diffuse away from
superbubbles, through the dense shell, without losing their energy
through coulombian losses, the LiBeB production induced by the SBEPs
outside the superbubbles is very small.  Any isolated supernova
exploding in the `unperturbed' ISM (i.e. far from SBs) then enriches
the ambient gas with freshly synthesized C and O without being
accompanied by an equivallent production of LiBeB. The gas around such
a SN can thus show very low L/M ratios, unless another mechanism
produces LiBeB in the same region.  Several processes can be invoked
for that purpose.  First, the standard GCRN: the shocks created by
isolated SNe accelerate CRs from the unperturbed ISM (mostly protons
and $\alpha$-particles) which then interact with the ambient CNO. The
ISM abundance of CNO being very low in the early Galaxy, the resulting
LiBeB production efficiency is much smaller than in SBs.  The
corresponding L/M production ratios are represented in Fig.~\ref{L/M}:
they increase linearly with metallicity, as expected for GCRN.

If this were the only production mechanism of light elements in the
unperturbed ISM, one should expect to find extremely low L/M ratios at
very low Z. However, we have shown in Parizot \& Drury (1999a,b) that
most of the metal-free CRs accelerated at the shock of an isolated SN
are actually confined inside the supernova remnant (SNR) during the
Sedov-like phase, and interact there with freshly ejected C and O
nuclei to produce significant amounts of Be and B. This means that
isolated SNe also produce LiBeB locally, where it is easily mixed with
the fresh CNO. We evaluated the production efficiency for this
mechanism to be about one order of magnitude lower than in
superbubbles.  The resulting L/M ratios are then about 10 to 30 times
below the most common values (obtained with the SB model), and should
be considered as a lower limit for L/M ratios in halo stars (provided
no depletion occurs after star formation, as can be checked from the
Li abundance).  This is represented by the lower horizontal line in
Fig.~\ref{L/M}.

\begin{figure}
\centerline{\psfig{file=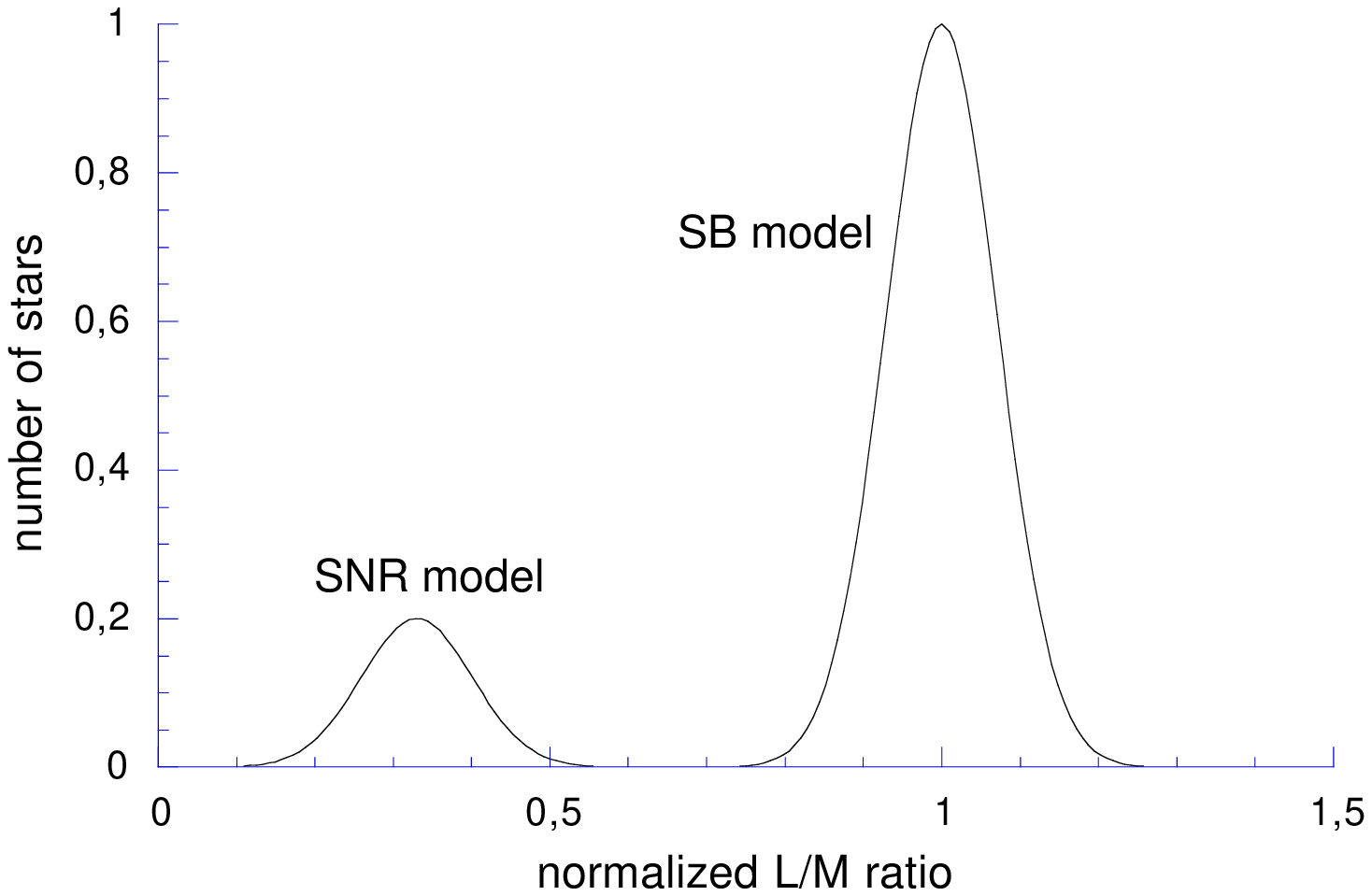, width=4.5cm}
            \hfill
            \psfig{file=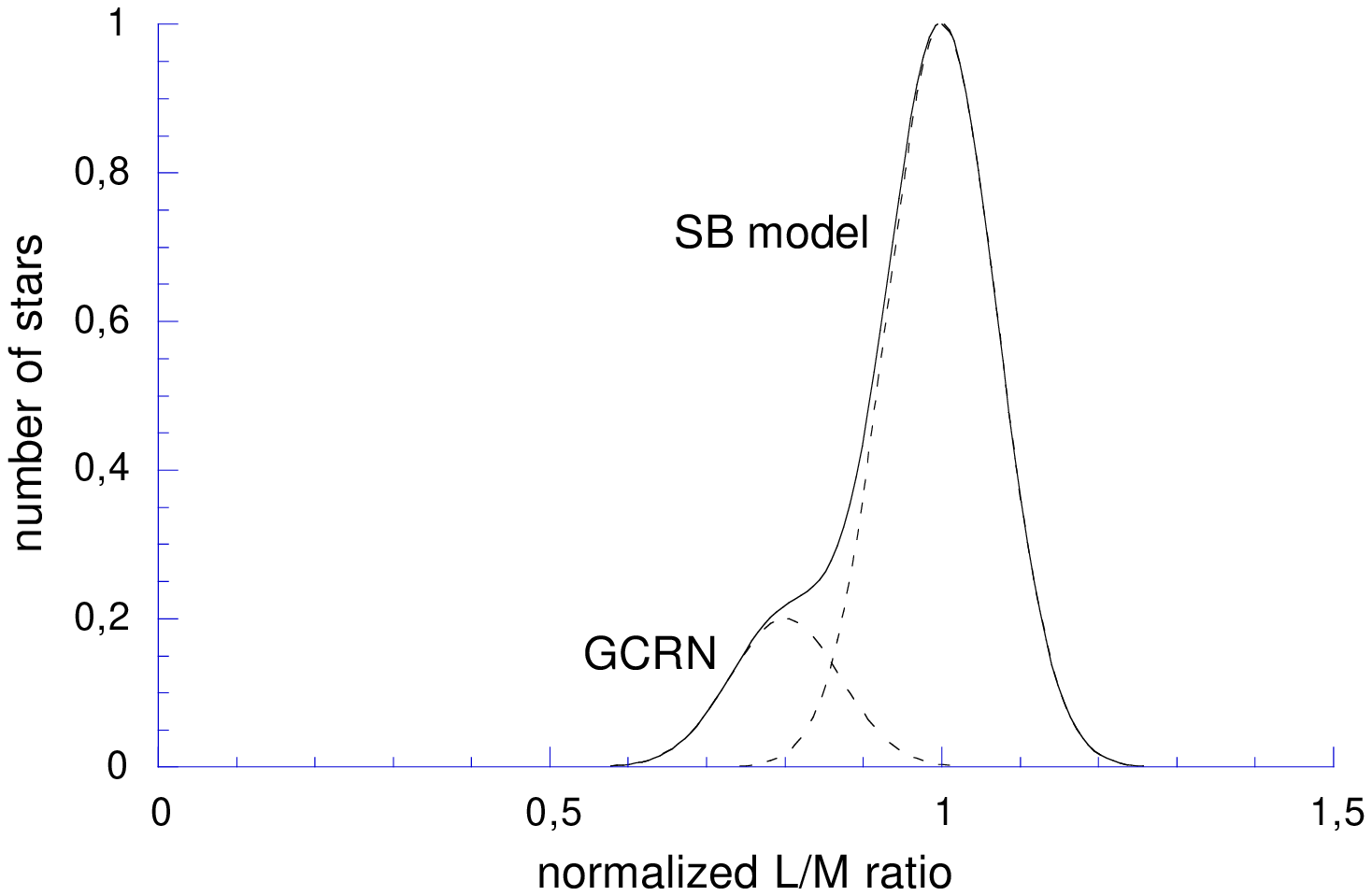, width=4.5cm}}
\caption{Schematic view of the histogram expected for the L/M ratios
in low-metallicity stars.  Left: very metal-poor stars,
$(\mathrm{O}/\mathrm{H}) \le 10^{-2}$ with the normalization of
Fig.~\ref{L/M}.  Right: intermediate metallicity, here (O/H) $= 0.8$.}
\label{bimodal}
\end{figure}

At very low metallicity, we thus predict a bimodal production of Be
and B, with SBEPs leading to a high efficiency mechanism (any of the
SB models) and CRs accelerated at the shock of isolated SNe leading to
a low efficiency mechanism (SNR model, Parizot \& Drury, 1999a,b).
This results in a bimodal distribution of the L/M ratios, as
schematically shown in Fig.~\ref{bimodal} (left).  Note that the
relative weight of the two `modes' depends on the fraction of stars
exploding in OB associations, and the fraction of stars forming far
from SBs.  At higher metallicity, when the Be and B production by GCRN
exceeds that of the SNR model, the distance between the peaks gets
smaller, and it is hard to distinguish between bimodality and the
scatter described in the previous section.  This is shown in
Fig.~\ref{bimodal} (right).

The ideal picture described above would be correct if there were no
mixing between the gas processed inside SBs (or their shells) and the
general ISM. In practice, this is only true during the first few
$10^{8}$ years of Galactic chemical evolution, when the Galaxy is
still largely inhomogeneous.  Later on, gas with high L/M ratios will
`pollute' the gas with low L/M ratios, leading to a broad L/M
distribution, rather than two distinct peaks.  Therefore, data at very
low-metallicity (say at $\mathrm{O}/\mathrm{H} \la 10^{-3}
(\mathrm{O}/\mathrm{H})_{\odot}$) are needed to fully test the model.
Most importantly, since the Li abundance in the early Galaxy is
dominated by the primordial $^{7}$Li, and thus unaffected by
spallative processes, only Be and B should be underabundant in the low
L/M stars.  The latter should thus show normal Li abundance, and
underabundant Be and B. Interestingly enough, Primas et al.  (1998)
reported such a behavior for the population~II star HD 160617, with a
deficiency of $\sim 0.5$~dex in B, at [Fe/H] $\sim -1.8$.  As recalled
by the authors, no stellar depletion process can be responsible for
the low B abundance observed, as any such process would deplete Li
much more than Be, because of its lower nuclear destruction
temperature.

We also wish to draw attention on the recent report by Boesgaard et
al.  (1999c) on two pairs of stars, (HD 84927, BD +20°3603) and (HD
94028, HD 219617), having the same stellar parameters (which limit the
risk of systematic errors in the derivation of the elemental
abundances) but Be abundances differing by as much as 0.3 and 0.6~dex,
respectively, at metallicities around [Fe/H] $\sim -2.1$ and $\sim
-1.5$ (or [O/H] $\sim -1.4$ and $\sim -0.85$).  This amounts to
``depletion'' factors of respectively 2 and 4.  It is still not clear
whether these differences are due to variations in the Be production
efficiency or to poor mixing of the SN ejecta with the gas containing
the secondary elements produced by spallation (cf.
Sect.\ref{scatter}).  Additional observations at lower metallicity
should allow us to draw more compelling conclusions in the next few
years.

\section{Conclusion}

We have shown that the SB models predict a scatter in the L/M ratios
observed in halo stars.  In the next few years, the statistical
analysis of this scatter (measured thanks to smaller error bars)
should provide information about the SB dynamics and the star
formation mechanism around SBs.  The accumulation of data should allow
us to distinguish between the two current SB models (good or poor
mixing of the gas inside and around SBs).

The models also predict a `bi-modal' production of Be and B in the
early Galaxy, with collective SNe giving rise to a high-efficiency
mechanism providing the observed L/M ratios (SB model), and isolated
SNe giving rise to a low-efficiency mechanism and L/M ratios 10 to 30
times lower (SNR model).  Both processes are local, respectively
inside superbubbles (or their shells) and inside SNRs, and independent
of the ambient ISM metallicity, as required by the observed or
inferred constancy of the L/M ratios at very-low metallicity.  In
addition to these processes, the standard GCRN is expected to occur on
the Galactic scale, but at a lower rate until the ISM metallicity
reaches about 3 to 10\% of the solar metallicity.  In any case, if two
populations of stars can be identified with respectively high and low
L/M ratios, the determination of their relative weight will give
information about the statistics of SN explosions in OB associations.

We also predict that Be will be found more deficient than B in the
so-called B-depleted stars, of which HD 160617 could only be a first
example.  On the other hand, if low-metallicity stars can be observed
with both strongly deficient Be \emph{and} B, with approximately the
same apparent ``depletion'', this would imply that the primary
component, i.e. the $\nu$-process, is not dominant for B, and that we
have to find an other process to account for the observed
$^{11}$B/$^{10}$B ratio.  This would put a strong constraint on light
element production, probably requiring the existence of very abundant
low-energy `cosmic-rays' ($E \sim 10-30$~MeV/n), powered by an energy
source still to be determined.

\begin{acknowledgements}
This work was supported by the TMR programme of the European Union
under contract FMRX-CT98-0168.
\end{acknowledgements}


\begin{thebibliography}{}

\bibitem[1999]{Boesgaard+99a} Boesgaard, A. M., King, J. R.,
Deliyannis, C. P., Vogt, S. S., 1999, AJ 117, 492

\bibitem[1999]{Boesgaard+99b} Boesgaard, A. M., Deliyannis, C. P.,
King, J. R., Ryan S. G., Vogt, S. S., Beers T. C., 1999, AJ 117, 492

\bibitem[1995]{Bykov95} Bykov, A. M., 1995, Space Sci. Rev. 74, 397

\bibitem[1999]{Bykov99} Bykov, A. M., 1999, in ``LiBeB, cosmic rays
and gamma-ray line astronomy'', R. Ramaty, E. Vangioni-Flam, M. Cass\'e,
K. Olive (eds.), ASP Conference Series, vol.  171, 146

\bibitem[2000]{Bykov+2000} Bykov A. M., Gustov M. Yu., Petrenko M. V.,
2000, in ??

\bibitem[1995]{Casse+95} Cass\'e, M., Lehoucq, R., Vangioni-Flam, E.,
1995, Nature 374, 337

\bibitem[1992]{Duncan+92} Duncan, D. K., Lambert, D. L., Lemke, M., 1992,
ApJ 401, 584

\bibitem{Duncan+97} D.K. Duncan, F. Primas, L.M. Rebull, A.M.
Boesgaard, C.P. Deliyannis, L.M. Hobbs, J.R. King and S.G. Ryan, ApJ
488 (1997) 338.

\bibitem[1999]{FieOli99} Fields B. D., Olive K. A., 1999, ApJ, 516, 797

\bibitem[2000]{Fields+2000} Fields B. D., Olive K. A., Vangioni-Flam
E., Cass\'e M., 2000, submitted to ApJ (astro-ph/9911320)

\bibitem[1999]{FulKra99} Fulbright, J. P., Kraft, R. P., 1999, AJ 118,
527

\bibitem[1998]{GarciaLopez+98} Garcia-L\'opez, R. J., Lambert, D. L.,
Edvardsson, B., Gustafsson, B., Kiselman, D., Rebolo, R., 1998, ApJ 500,
241

\bibitem[1998]{Higdon+98} Higdon, J. C., Lingenfelter, R. E., Ramaty, R.,
1998, ApJ 509, L33

\bibitem[1998]{Israelian+98} Israelian, G., Garc\'ia-L\'opez, R. J.,
Rebolo, R., 1998, ApJ 507, 805

\bibitem[1971]{Meneguzzi+71} Meneguzzi, M., Audouze, J., Reeves, H.,
1971, A\&A 15, 337

\bibitem[1997]{Molaro+97} Molaro, P., Bonifacio, P., Castelli, F.,
Pasquini, L., 1997, A\&A 319, 593

\bibitem[1999]{ParDru99a} Parizot, E., Drury, L., 1999a, A\&A 346, 329

\bibitem[1999]{ParDru99b} Parizot, E., Drury, L., 1999b, A\&A 346, 686

\bibitem[1999]{ParDru99c} Parizot, E., Drury, L., 1999c, A\&A 349, 673

\bibitem[2000]{ParDru2000} Parizot, E., Drury, L., 2000, submitted to
A\&A

\bibitem[1998]{Primas+98} Primas, F., Duncan, D. K., Thorburn, J. A.,
1998, ApJ 506, L51

\bibitem[1999]{Ramaty+99} Ramaty, R., Scully, S. T., Lingenfelter, R. E.,
Kozlovsky, B., 1999, ApJ submitted, astro-ph/9909021

\bibitem[1970]{Reeves+70} Reeves, H., Fowler, W. A., Hoyle, F., 1970,
Nature 226, 727

\bibitem[1990]{Vangioni+90} Vangioni-Flam, E., Cass\'e, M., Audouze,
J., Oberto, Y., 1990, ApJ 364, 586.

\bibitem[1990]{Woosley+90} Woosley, S. E., Hartmann, D. H., Hoffman,
R. D., Haxton, W. C., 1990, ApJ 356, 272

\end{thebibliography}
\end{document}